\documentclass{PoS}
\usepackage{amsmath,slashed}
\newcommand\Tstrut{\rule{0pt}{2.6ex}} 

\usepackage[
    style=numeric-comp,
    backend = bibtex,
    bibencoding = ascii,    
    sorting=none,
    firstinits=true,
    isbn=false,
    url=false,
    doi=false,
    maxbibnames = 99
    ]{biblatex} 
    
\DeclareFieldFormat
[article,inbook,incollection,inproceedings,patent,thesis,unpublished,suppbook,suppcollection,suppperiodical]
  {title}{\mkbibemph{#1}}
\DeclareFieldFormat{eprint:arxiv}{[#1]}

\renewbibmacro{in:}{} 

\title{Spectroscopy of two dimensional N=2 Super Yang Mills theory}

\ShortTitle{Spectroscopy of two dimensional N=2 Super Yang Mills theory}

\author{\speaker{Daniel August}%
        \\
     Theoretisch-Physikalisches Institut, Friedrich-Schiller-Universit\"at Jena, 07743 Jena, Germany\\
       E-mail: \email{daniel.august@uni-jena.de}}

\author{Bj\"orn Wellegehausen\\
		Institut f\"ur Theoretische Physik, Justus-Liebig-Universit\"at Giessen, 35392 Giessen, Germany
and \\
Theoretisch-Physikalisches Institut, Friedrich-Schiller-Universit\"at Jena, 07743 Jena, Germany\\
       E-mail: \email{bjoern.wellegehausen@uni-jena.de}}

\author{Andreas Wipf\\
       Theoretisch-Physikalisches Institut, Friedrich-Schiller-Universit\"at Jena, 07743 Jena, Germany\\
       E-mail: \email{wipf@tpi.uni-jena.de}}

\abstract{Albeit the standard model is the most successful model of particles physics, it still has some theoretical shortcomings, for instance the hierarchy problem, the absence of dark matter, etc. Supersymmetric extensions of the standard model could be a possible solution to these problems.\\

One of the building blocks of these supersymmetric models are supersymmetric gauge theories. It is expected that they exhibit interesting features like confinement, chiral symmetry breaking, magnetic monopoles and the like.\\ We present new results on N=2 Super Yang Mills theory in two dimensions. The lattice action is derived by a dimensional reduction of the N=1 Super Yang Mills theory in four dimensions. By preserving the R symmetry of the four dimensional model we can exploit Ward identities to fine tune our parameters of the model to obtain the chiral and supersymmetric continuum limit. This allows us to calculate the mass spectrum at the physical point and compare these results with effective field theories.}

\FullConference{34th annual International Symposium on Lattice Field Theory\\
         24-30 July 2016\\
         University of Southampton, UK}

\begin{document}

\section{Introduction}

Supersymmetric gauge theories are the building blocks of a supersymmetric standard model. One of these theories is the $N=1$ Super Yang Mills (SYM) theory in four dimensions featuring a gauge field and a Majorana fermion. Although this model is most likely not realized in nature, we may learn many interesting features of supersymmetric gauge theories by investigating this model.\newline
Here we will study a dimensionally reduced version of the theory, namely $N=2$ SYM theory in two dimensions. This model attracted a lot of interest in the recent years because it allows for a lattice formulation which preserves a part of supersymmerty on the lattice exactly \cite{CatKapUns2009,BerCat2016}.\newline
In contrast to this exact formalism we will use the conventional 
lattice formulation close to the proposal put forward in \cite{SuzTan2005}.  

\section{N=2 SYM in two dimensions}

Our model is described by the action

\begin{multline}
	S=\frac{1}{g^2}\int d^2 x \textrm{ tr}\left\{-\frac{1}{4} F_{\mu\nu}F^{\mu\nu} + \frac{i}{2}\overline{\chi}_r \Gamma_\mu D^\mu \chi^r-\frac{1}{2} D_\mu\phi_a D^\mu\phi^a \right. \\ \left. + \frac{i}{2} \overline{\chi}_r \left(\sigma_1\right)^r_s \gamma_5\left[\phi^1,\chi^s\right] + \frac{i}{2} \overline{\chi}_r \left(\sigma_3\right)^r_s \gamma_5 \left[\phi^2,\chi^s\right] + \frac{1}{4}\left[\phi_a,\phi_b\right]\left[\phi^a,\phi^b\right]\right\}. \label{PoS1}
\end{multline}
where $g$ is the two dimensional coupling constant, $F_{\mu\nu}$ is the field strength tensor, $\chi_r$ are Majorana fermions, $\Gamma_\mu$ are the two dimensional gamma matrices, $\phi_r$ are real scalars and $\sigma_i$ are the Pauli matrices acting in
flavour space. The fermions and the scalars are in the adjoint representation of the gauge group, which we have chosen to be $SU(2)$.\\
The action possesses the two dimensional Lorentz symmetry, a chiral symmetry and a $SO(2)$ flavour symmetry. One important point is that the scalar potential $[\phi_1,\phi_2]^2$ possesses flat direction. These could spoil a lattice simulation if they do not get lifted by quantum effects.\newline
We use the tree-level Symanzik improved gauge action and Wilson fermions to construct the lattice formulation.

\section{Ward Identities}

The theory is super renormalizable and in the lattice formulation has only one relevant operator $m^2_s\phi^2$, giving us the scalar bare mass $m_s$ as a parameter to construct a supersymmetric continuum limit \cite{SuzTan2005}. The 
only problem left is to find an appropriate operator which allows for this fine-tuning. 
We shall use Ward Identities following from the supersymmetry (susy) variation of a 
suitably chosen term $T$.

We begin with recalling the two susy transformations of the fields,
\begin{align}
	q_1\chi_1&=\gamma_{\mu\nu}F^{\mu\nu}+i\gamma_\mu\gamma_5 D^\mu \phi_2 & 
	q_2\chi_1&=i\gamma_\mu\gamma_5 D_\mu \phi_1+i\left[\phi_1,\phi_2\right] \nonumber \\
	q_1\chi_2&=i\gamma_\mu\gamma_5 D_\mu \phi_1-i\left[\phi_1,\phi_2\right] &
	q_2\chi_2&=\gamma_{\mu\nu}F^{\mu\nu}-i\gamma_\mu\gamma_5 D^\mu \phi_2 \nonumber \\
	q_1 A_\mu &= \frac{1}{2}\overline{\chi}_1 \gamma_\mu & 
	q_2 A_\mu &= \frac{1}{2}\overline{\chi}_2\gamma_\mu \nonumber \\
	q_1 \phi_1&=\frac{i}{2}\overline{\chi}_2\gamma_5 & 
	q_2 \phi_1&=\frac{i}{2}\overline{\chi}_1\gamma_5 \nonumber \\
	q_1 \phi_2&=\frac{i}{2}\overline{\chi}_1\gamma_5 &
	q_2 \phi_2&=-\frac{i}{2}\overline{\chi}_2\gamma_5. \label{PoS2}
\end{align}
The cleverly chosen term
\begin{equation}
	\left(T(x)\right)_a=g\, \mathrm{tr}_c\left\{\overline{\chi}_b(x){\left(\sigma_1\sigma_3\right)^b}_a [\phi_1,\phi_2] (x)\right\}, \label{PoS3}
\end{equation}
leads to  a Ward Identity relating the scalar potential and the Yukawa terms
\begin{equation}
	\left<\frac{g^2}{2}\left[\phi_1(x),\phi_2(x)\right]^2\right>=
	\frac{g}{4}\left<\frac{i}{2}\overline{\lambda}(x)\Gamma_2\left[\phi_1,\lambda(x)\right]+\frac{i}{2}\overline{\lambda}(x)\Gamma_3\left[\phi_2,\lambda(x)\right]\right> \,.\label{PoS4}
\end{equation}
We use this Ward identity as it is most sensitive in the 
parameter range relevant for fine-tuning.

\section{Mass Spectrum}
We expect the particle spectrum given in Table \ref{Tabel1}, in which
we identified the particles by their quantum numbers.

\begin{table}[h!]
\begin{center}
\begin{tabular}{c|c|c}
	particle & spin & name \\ \hline
	$\overline{\chi}\gamma_5\chi$ & 0  & a-$\eta$' \Tstrut \\
	$\overline{\chi}\chi$ & 0  & a-$f_0$ \\
	$F_{\mu\nu}\Sigma^{\mu\nu}\lambda$ & $\frac{1}{2}$ & gluino-glueball
\end{tabular}
\end{center}
\begin{center}
\begin{tabular}{c|c|c}
	particle & spin & name \\ \hline
	$[\phi_1,\phi_2]F_{\mu\nu}$ & 0 & glue scalarball  \Tstrut \\
	$F_{\mu\nu}F^{\mu\nu}$,$[\phi_1,\phi_2]^2$ & 0 & $0^+$-glueball, scalarball\\
	$F_{\mu\nu}\Sigma^{\mu\nu}\chi$,$[\phi_1,\phi_2] \sigma_2\chi$ & $\frac{1}{2}$ & gluino-glueball, gluino-scalarball
\end{tabular} 
\end{center}
\caption{Two dimensional reduced supermultiplets for the theory}
\label{Tabel1}
\end{table}
Another way to observe the restoration of susy is to study the spectrum 
of the theory since in the supersymmetric limit particles in the same multiplet
have equal mass.
Using the results of \cite{VenYan1982,FarGabSch1998} we find 
the two dimensional reduced supermultiplets given in Tabel \ref{Tabel1}.

%

\section{Results}

\subsection{Flat directions}

\begin{figure}
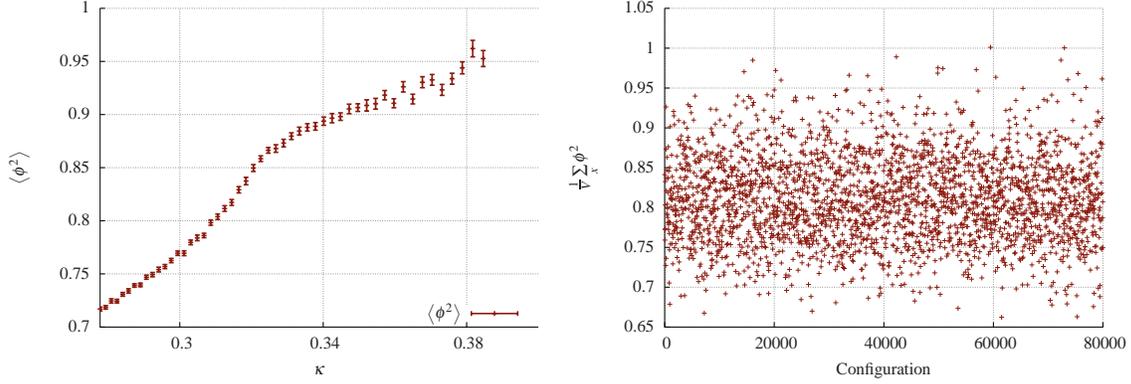

\begin{minipage}[b]{0.49\textwidth}
\resizebox{!}{0.25\paperwidth}
{
	\input{Flatdirections.tex}
}
\end{minipage}
\begin{minipage}[b]{0.49\textwidth}
\resizebox{!}{0.25\paperwidth}
{
	\input{FlatdirectionsOnConfig.tex}
}
\end{minipage}
\caption{On the left: value of $\left<\phi^2\right>$ depending 
on the hopping parameter $\kappa$ for $m_s=0$. On the right: 
spatial average of $\phi^2$ for $\kappa$ = 0.3125, chiral limit: $\kappa_\mathrm{crit}\approx 0.31$}
\label{Fig1}
\end{figure}

First of all we consider the problem with flat directions in the scalar
sector of the model.
Fig. \ref{Fig1} shows the expectation value of $\phi^2$ for different 
values of the hopping parameter $\kappa$ and for $m_s=0$. Since a positive scalar mass should stabilize the flat directions, the $m_s=0$ case should have the biggest problems. 
But as we see, even in this case the expectation value stays finite and as shown on the right hand side, we observe no exponential growth associated with flat directions. In the lattice regularization the flat directions are stabilized.
Therefore we should not encounter any problems in our simulations at 
$m_s>0$. Indeed, monitoring $\phi^2$, we did not 
meet any instabilities.

\subsection{Ward Identities}

\begin{figure}
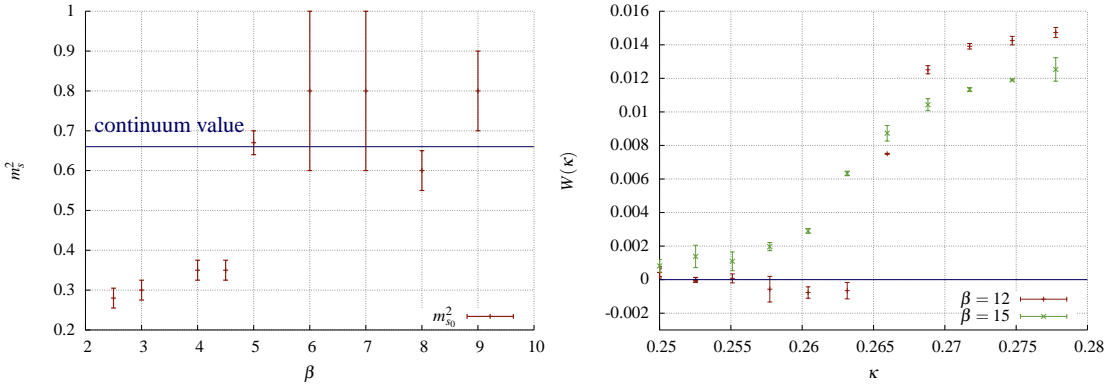

\begin{minipage}[t]{0.48\textwidth}
\resizebox{!}{0.25\paperwidth}{\input{Msdependace.tex}}
\end{minipage}
\begin{minipage}[t]{0.48\textwidth}
\resizebox{!}{0.25\paperwidth}{\input{ComparisonWard.tex}}
\end{minipage}
\caption{Left: Plot of the value $m^2_{s_0}$ for which the Ward identity is satisfied for different $\beta\equiv 1/g^2$. Right: Plot of the Ward Identity for fixed $m^2_s=0.66$ dependent on $\kappa$ for two different values of $\beta$.}
\label{Fig2}
\end{figure}
Defining $m_{s_0}$ as the mass for which the Ward identity is satisfied, we can see in Fig. \ref{Fig2} the dependence of $m^2_{s_0}$ on $\beta$, where we have chosen $\kappa(\beta)$ in such a way, that we are close to the chiral limit. We observe two regions of $\beta$ values, first $\beta<5$ where the value $m^2_{s_0}$ is in the range $0.3-0.35$ which is approximately half of the expected continuum value \cite{SuzTan2005}. 
For $\beta\geq 5$ the value jumps to $0.6-1.0$ with large error bars.
We conclude that for $\beta$-values below $5$, we are in a phase irrelevant
for the supersymmetric continuum limit.\newline
On the right hand side of Fig. \ref{Fig2} we see the $\kappa$-dependence of the Ward identity for two different $\beta$ and the continuum value $m^2_s=0.66$. 
For $\beta=15$ the Ward identity is not satisfied. This could be a lattice artefact for large $\beta$, similarly as the zero crossing shown for $\beta=12$. Therefore we use a different approach for choosing the scalar mass. We use the continuum value of $m_s^2=0.66$ while still fine-tuning $\kappa$. This way we obtain sensible and accurate 
results in the meson sector.

\subsection{Meson Spectrum}
\begin{figure}
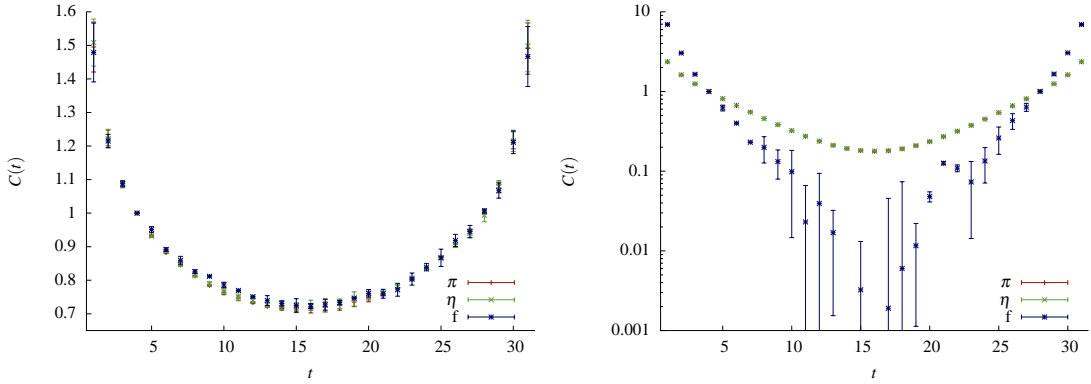

\begin{minipage}[t]{0.48\textwidth}
\resizebox{!}{0.25\paperwidth}{\input{Mesons12.tex}}
\end{minipage}
\begin{minipage}[t]{0.48\textwidth}
\resizebox{!}{0.25\paperwidth}{\input{Mesons08.tex}}
\end{minipage}
\caption{Meson correlators for the critical $\kappa=0.266$ (left) 
and $\kappa=0.260$ (right) for $\beta=12$ and $m^2_s=0.66$.}
\label{Fig3}
\end{figure}

Fig. \ref{Fig3} shows the meson correlators for two different $\kappa$ and $\beta=12$. The first $\kappa$ is the critical $\kappa=0.266$ for which the chiral susceptibility has a peak. We observe that all three correlators are degenerate which points to a restoration of susy. The other value $\kappa=0.260$ is close to the critical value.
\begin{figure}
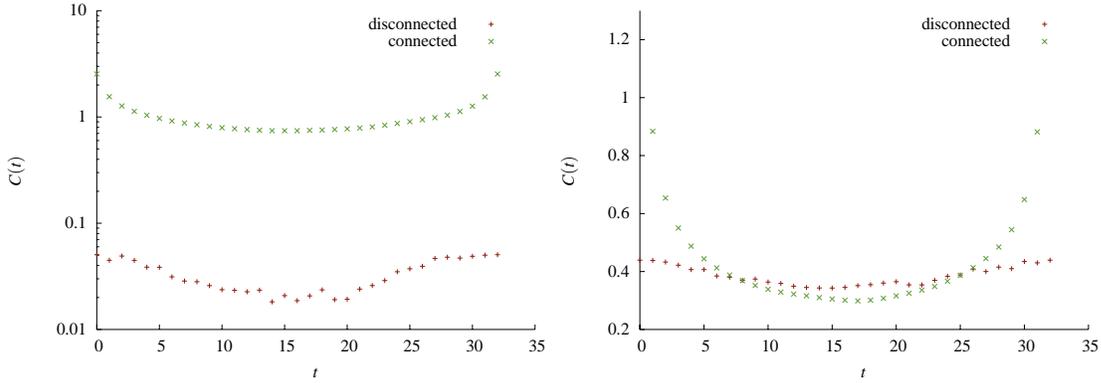

\begin{minipage}[t]{0.48\textwidth}
 \resizebox{!}{0.25\paperwidth}
{\input{etaComponents.tex}}
 \end{minipage}
  \begin{minipage}[t]{0.48\textwidth}
 \resizebox{!}{0.25\paperwidth}
{\input{fComponents.tex}}
 \end{minipage}
 \caption{Connected and disconnected part of the $\eta$ (left) and $f$ (right) correlator for $\beta=12$ and $\kappa=0.266$.}
 \label{Fig4}
\end{figure}
Here the 
$\eta$ and the $\pi$ mesons are degenerate while the $f$ meson is not. A preliminary
analysis of the masses show that the excited $\pi$ and $\eta$ mesons are 
in fact heavier than the possible excited state of the $f$ meson. But the contribution 
to their correlators is orders of magnitudes larger for the $f$ meson, 
leading to a steeper slope of the $f$ correlator.
\newline
Coming back to the degeneracy of the three states we are especially interested in the $\eta$ and $f$ meson, since they appear in the supermultiplet (see Table \ref{Tabel1}). 
Thus we look separately at their disconnected and connected parts, as plotted in Fig. \ref{Fig4}. We see that the disconnected part of the $\eta$ is smaller by a factor of $10-100$. Since the $\pi$ correlator is the connected part of the $\eta$ correlator,  this explains why both states are degenerate. On the other hand the disconnected and connected parts of the $f$ meson have comparable magnitudes and we have no easy explanation for the degeneracy of the $\eta$ and $f$. Therefore this degeneracy seems to be a non 
trivial test for the restoration of susy.\newline
If we restore susy for $\kappa\to 0.266$ and $\beta=12$ we observe two effects: As
in four dimensions we restore susy in the chiral limit \cite{CurVen1987,Suz2012}
and second the $\pi$ becomes massless in this limit. Thus the $\eta$ and $f$
should become massless if the degeneracy with the $\pi$ meson holds in the continuum limit.

\subsection{Glueball}
Fig. \ref{Fig5} shows the glue ball correlator in the pure gauge theory and the $N=2$ SYM in two dimensions for different smearing. 
Here we define $S$ as the product of the smearing steps and the smearing parameter.
The first observation is that the slope of the correlators bends in the wrong
direction.
\begin{figure}
\begin{minipage}[t]{0.48\textwidth}
 \resizebox{!}{0.25\paperwidth}
{\input{GaugeTheory.tex}}
 \end{minipage}
  \begin{minipage}[t]{0.48\textwidth}
 \resizebox{!}{0.25\paperwidth}
{\input{GlueBallCorr.tex}}
 \end{minipage}
 \caption{Correlator for the glueball in the purge gauge theory(left) and the two dimensional $SYM$ theory(right) for different smearings S(=steps$\cdot$parameter)}
 \label{Fig5}
\end{figure}
\begin{figure}
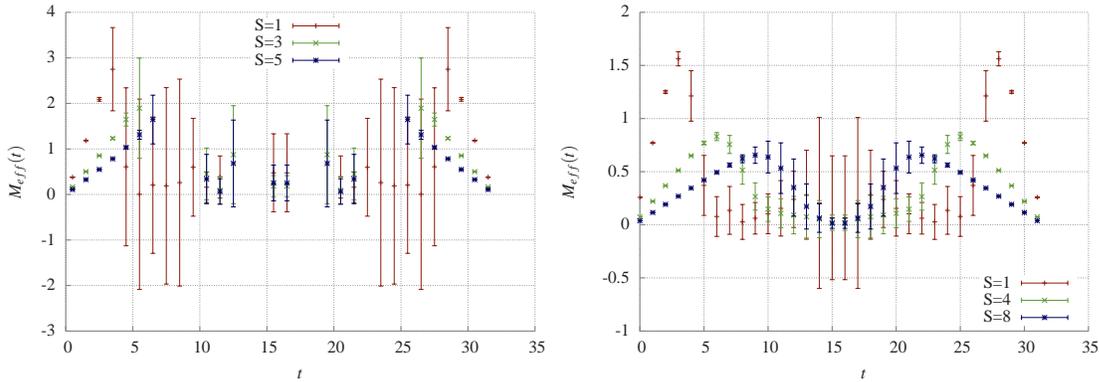

\begin{minipage}[b]{0.48\textwidth}
 \resizebox{!}{0.25\paperwidth}
{\input{GaugeTheoryMass.tex}}
 \end{minipage}
 \begin{minipage}[b]{0.48\textwidth}
 \resizebox{!}{0.25\paperwidth}
{\input{GlueBallEffMass.tex}}
\end{minipage}
\caption{Effective mass for the glueball in the pure gauge theory (left) and the two dimensional $SYM$ theory (right) for different smearings $S$(=steps$\cdot$parameter)}
 \label{Fig6}
\end{figure}
This is true for all three levels of smearing. Second the correlator in the 
pure gauge theory is constant after it reaches the value $0.001$. 
To understand this feature we calculated analytically the glue ball correlator for the two dimensional pure gauge theory using Migdals description \cite{Mig1975}. If we call $G(x)$ the glue ball operator which is formed by an arbitrary set of link enclosing the surface $A_G$ and has the right quantum numbers we find
\begin{equation}
\left<G(x)G(y)\right>=C_G=\mathrm{const.}\hskip6mm\hbox{ if }A_{G(x)}
\hbox{ and }  A_{G(y)}
\hbox{ do not overlap or touch each other.} \label{PoS6}
\end{equation}
Choosing for $G(x)$ the plaquette variable one can show that $C_G$ vanishes in the continuum limit. Therefore the glue ball decouples from the theory or an other interpretation is that the mass of the glue ball becomes infinity. This coincides with the analytical results presented  
in \cite{Bra1980}.\newline
With this result we can understand the correlator depicted in Fig. \ref{Fig5}: it 
turns into a constant if the time-separation  becomes large enough. In Fig. \ref{Fig6} we show the effective mass for both correlators. Both figures show very similar behaviour. 
Thus we conclude that the glue ball in $N=2$ SYM is infinitely heavy too.

\acknowledgments

D.A. and B.W. were supported by the DFG Research Training Group 1523/2 ``Quantum and Gravitational Fields ''. B.W. was supported by the Helmholtz International Center for FAIR within the LOEWE initiative of the State of Hesse.


\printbibliography 

\end{document}